\documentclass{JHEP3}
\usepackage{amsmath}
\usepackage[numbers,sort&compress]{natbib}
\usepackage{graphicx}
\usepackage{url}
\usepackage[labelfont=bf,font=small,width=.9\textwidth]{caption}
\makeatletter
\renewcommand\@ENVwarn[1]{}
\makeatother

\title{Temperature versus acceleration: the Unruh effect for holographic models}

\author{Angel Paredes{}$^1$, Kasper Peeters{}$^{1,2}$ and Marija Zamaklar{}$^2$\\
\llap{{}$^1$}Institute for Theoretical Physics, Utrecht University, P.O.~Box 80.195,
3508 TD Utrecht, The Netherlands.\\
\llap{{}$^2$}Department of Mathematical Sciences,
Durham University,
South Road,
Durham DH1 3LE, United Kingdom.\\
~\\
\email{a.paredesgalan@uu.nl}\\
\email{kasper.peeters@aei.mpg.de}\\
\email{marija.zamaklar@durham.ac.uk}}

\abstract{We analyse the effect of velocity and acceleration on the
  temperature felt by particles and strings in backgrounds relevant in
  holographic models. First, we compare accelerated strings and
  strings at finite temperature.  We find that for fixed Unruh
  temperature felt by the string endpoints, the screening length is
  smaller for the accelerated Wilson loop than for the static one in a
  thermal background of the same temperature; hence acceleration
  provides a ``more efficient'' mechanism for melting of
  mesons. Secondly, we show that the velocity-dependence of the
  screening length of the colour force, previously obtained from a
  moving Wilson loop in a finite temperature background, is not
  specific for the string, but is a consequence of the generic fact
  that an observer which moves with constant velocity in a black hole
  background measures a velocity-dependent temperature.  Finally, we
  analyse accelerated particles and strings in the AdS black hole
  background, and show that these feel a temperature which increases
  as a function of time. As a byproduct of our analysis we find a
  global Minkowski embedding for the planar AdS black hole.}

\preprint{\small ITP-UU-08/71, SPIN-08/54, DCPT-08/65}

\begin{document}
%====================================================================
\section{Introduction and summary}

Evidence is mounting that there are various interesting phenomena
related to the behaviour of mesonic resonances at finite temperature,
just above the QCD deconfinement transition. Lattice studies~(see
e.g.~\cite{Aarts:2007pk}) have shown that heavy quark
bound states can survive rather deep into the deconfined phase, and
various experimental ideas are being pursued to measure such states
directly and disentangle their decay
channels~\cite{Oda:2008kg}. However, when the bound states move or
accelerate with respect to the surrounding plasma, a lattice analysis
becomes impossible (at least with current techniques) and other
approaches are required.

Holographic methods, based on the string/gauge theory correspondence,
have proven useful to analyse such dynamical situations. The
correspondence maps the behaviour of strongly coupled gauge theories
at finite temperature to physics of weakly coupled strings in black
brane backgrounds. The gauge theories for which a string dual is known
are so far unfortunately only caricatures of QCD. However, the
qualitative kinematical behaviour of quark-antiquark bound states
obtained with this method shows interesting similarities with results
from lattice gauge theory. In confining chiral models such as those
of~\cite{Sakai:2004cn} there indeed exists a phase in which there are
still quark-antiquark resonances when the gluons have already
deconfined, and e.g.~the behaviour of their thermal masses is similar
to that obtained from the lattice~\cite{Peeters:2006iu}.

The main power of holographic methods, however, is that they can
provide insight into the physics of more dynamical situations, in
which quarks and mesons do not sit at rest with respect to the
surrounding plasma. In contrast to lattice simulations, holographic
methods do not require a Wick rotation to the Euclidean regime, and
deal with finite temperature directly in real time. A first result in
this direction was the computation of the drag force on free
quarks~\cite{Herzog:2006gh,Gubser:2006bz}. Such a drag force is not
present for colour singlet states, but they do exhibit other
interesting
effects. In~\cite{Peeters:2006iu,Liu:2006nn,Chernicoff:2006hi} it was
shown that the temperature at which mesons dissociate decreases with
increasing velocity, i.e.~with increasing transverse momentum. In more
detail, it was shown that the screening length of the colour force
decreases with the velocity~$v$ of the quark-antiquark bound state
as~$L \sim (1-v^2)^{1/4}/T$, where~$T$ is the temperature. This result
has turned out to be rather robust against changes of the string/gauge
theory dual, and a similar relation also holds for low-spin
mesons~\cite{Ejaz:2007hg} and
baryons~\cite{Athanasiou:2008pz,Krishnan:2008rs}. Interpreted the
other way around, it states that dissociation of a colour singlet is
inevitable for sufficiently large velocity (or transverse momentum).
\medskip

Given the usefulness of holographic models to study finite-temperature
physics, and given the relation of temperature to acceleration through
the Unruh effect~\cite{Unruh:1976db}, it is natural to ask whether
results such as those described above have any analog for accelerated
colour singlets. This question was analysed in~\cite{Peeters:2007ti},
where it was shown that an interesting bound also exists on the
acceleration of mesons at zero temperature. A simple relation between
the maximal acceleration and the angular momentum of a meson was
obtained, $a_{\text{max}} \sim \sqrt{T_s/J}$, where~$T_s$ is the
string tension. This result holds for strings in flat space,
accelerating orthogonally to the angular momentum plane. At a
technical level the computation is closely related to the computation
of the angular momentum bound for holographic mesons at finite
temperature~\cite{Peeters:2006iu}.

As the computation of~\cite{Peeters:2007ti} was done only in flat
space, the next step is obviously to combine finite-temperature
backgrounds with accelerated probes. This is the topic of the present
paper. We will analyse a variety of strings and particles that move
with a constant velocity or acceleration in empty AdS space or in a
finite-temperature holographic background. One would naively expect
that an accelerated particle or string in a finite-temperature
holographic background is subject to an effective temperature that is
a ``superposition'' of the background temperature and the Hawking
temperature due to the acceleration. The situation, however, turns out
to be more subtle.  For instance, while in flat space all particles
that move with constant velocity feel the same (vanishing)
temperature, this is no longer the case at non-zero temperature, as we
will show below.  The situation becomes even more subtle when
acceleration effects are combined with a finite-temperature
background.

In most physical situations, accelerations are such
that the Unruh temperature is negligible. On the other hand, it is
natural to expect that strong interactions produce the largest
possible accelerations. An example is jet quenching inside a
quark-gluon plasma, which shows that highly energetic partons are
stopped in just a few fermi. It would be of great interest to
understand if the large Unruh temperature that such partons would
experience could have any phenomenological consequence. For a proposal
in which Unruh temperature plays a relevant role for hadronic physics
in a different situation, see e.g.~\cite{Castorina:2007eb}.

\medskip

The outcome of our analysis is interesting in a number of ways.  In the
first part of this paper, we analyse various moving and accelerated
probes in an AdS background, which are generalisations of the
configurations studied in~\cite{Peeters:2007ti}. As expected, the
qualitative results of~\cite{Peeters:2007ti} persist. However, when
comparing the accelerated string configurations at zero temperature
with finite temperature static string configurations, we observe that
acceleration is in general a more efficient method to dissociate
colour singlets than temperature (in a way which will be made precise
later).

Perhaps the most striking result is obtained in the second part of the
paper, where we turn to finite-temperature situations. This requires
us to analyse how the Hawking temperature changes when it is observed
by a moving or accelerated observer.\footnote{How temperature behaves
  under Lorentz boosts is an issue that has led to a lot of
  controversy in the last century; we will comment on this further in
  \S\ref{s:va_finite_T}.}  It turns out that even without knowing the full
evolution of an accelerated string, one can already draw strong
conclusions simply based on the effective temperature felt by its
centre of mass. In fact, we will show that even the velocity
dependence of the screening length, \mbox{$L \sim (1-v^2)^{1/4}/T$ }is a
simple consequence of the observer dependence of temperature. For
accelerated motion, we will argue that the temperature rises
monotonically as a function of time.

%Entropy of two-horizon system:~\cite{Shankaranarayanan:2003ya}.

%====================================================================
\section{Temperature effects due to acceleration in curved space}

In the first part of this paper we will consider temperature effects
that are caused by acceleration with respect to an inertial observer
in curved backgrounds. The fact that our background possesses Lorentz
symmetry in the directions parallel to the boundary guarantees that
any two observers which move with constant 4-velocity in directions
parallel to the boundary will not accelerate with respect to each
other, and will hence measure the same temperature. We will see later
that this ceases to be the case when a black hole is put in the
background.

In general, an arbitrarily moving observer will not be in a (local)
thermal bath, and whether it is possible to define a temperature for
such an observer, and which value of the temperature it measures,
crucially depends on the space-time path of the observer. There are
different ways in which one may try to determine the temperature which
is felt by the observer. One way is to model quantum-mechanically the
detector carried by the observer, and use quantum field theory to
determine the spectrum which it measures (see
e.g.~\cite{birr1}). However, this method requires at least access to
the Wightman function for a scalar field on the curved background. For
most backgrounds we consider in this paper, the Wightman function is
not available in closed form. We will therefore employ more
geometrical approaches to the problem: the surface gravity method and
Global Embedding in Minkowskian Space-time method (the GEMS), which
obtains an effective temperature for a path in a curved background by
embedding the path in a higher-dimensional flat space.  Both
techniques have been used to find the temperature observed by various
types of observers in a variety of backgrounds, including (A)dS,
Schwarzschild and Reissner-Nordstrom~\cite{Deser:1998xb}. Before
applying these techniques, we will first revise some of the basic
ideas behind them.

%--------------------------------------------------------------------
\subsection{The surface gravity and GEMS computation of temperature}

A long time ago it was shown by Unruh~\cite{Unruh:1976db} that an
observer which moves on a \emph{straight} line with constant
acceleration in \emph{flat} space-time sees a vacuum filled with a
thermal distribution of particles, and measures a temperature which is
proportional to its acceleration
\begin{equation}
\label{accelT}
T_U = \frac{1}{2 \pi} a \, .
\end{equation}
This however is not the case for a generic path in flat
or curved space, on which a detector will typically not detect a
thermal spectrum~\cite{Letaw:1979wy}.  However, for those special orbits for
which a temperature can be defined, there are two simple geometrical approaches
which can be used to determine the temperature.  These are the method
of surface gravity and the method which uses global embedding in
a (higher-dimensional) Minkowskian space-time (the GEMS method).  It
turns out that there are also situations when one of the methods can
be used but not the other, and we will discuss such cases in~\S\ref{s:particle_accel_AdSBH}.

The surface gravity approach claims that an observer following an
orbit of a time-like Killing vector field measures a temperature
given by Tolman's law,
\begin{equation}
\label{TL}
T = \frac{k_H}{2\pi\,\sqrt{\varsigma^2}}\,,
\end{equation}
where~$\varsigma^\mu$ denotes the Killing vector and $k_H$ is the
surface gravity of the horizon. The surface gravity in turn is defined
by
\begin{equation}
\label{kH}
k_H^2 = - \frac{1}{2}(\nabla^\mu \varsigma^\nu)(\nabla_\mu \varsigma_\nu)\bigg{|}_{\text{at the horizon}} \, . 
\end{equation}
In this formula the horizon can be observer-dependent (e.g.~it is the
Rindler horizon for an accelerated particle in flat
space).\footnote{We note that there is freedom in the normalisation of
  $\varsigma$ at infinity; generically one is allowed to multiply
  $\varsigma$ with a constant, which preserves the property that it is
  a Killing vector. However, it is clear from~\eqref{TL} that such a
  freedom does not affect the measured temperature as obtained from
  Tolman's law.}

The alternative GEMS approach builds on the fact that any curved
Lorentzian space can be embedded in some higher-dimensional flat
Minkowski space, and that hence any motion in an arbitrary curved
space can be equivalently described by considering motion of the
particle in that higher-dimensional flat Minkowski space. If it turns
out that the higher-dimensional orbit is a time-like orbit of constant
higher-dimensional acceleration, then the standard Unruh arguments can
be used to define the temperature. In these cases the
higher-dimensional Unruh temperature takes the form of Tolman's law,
and agrees with the surface gravity computation~\eqref{TL}.  If
however, the GEMS orbit turns out to be ``far away'' from a constant
acceleration orbit, the GEMS approach cannot be used. This can happen
\emph{even} when the temperature in the original system shows no
pathologies, so one has to be careful.

A quantitative measure of how close the local Unruh temperature is to
the temperature obtained from Tolman's law, i.e.~how slow the rate of
``change'' of the temperature is, was proposed
in~\cite{Russo:2008gb}. This involves generalising the notion of
``jerk'', being the rate of change of acceleration in classical
non-relativistic mechanics. A relativistic jerk was defined as
\begin{equation}
\Sigma^k = j^k - a^2 v^k,  \qquad j^k = v^i D_i a^k \, , 
\end{equation}
where the indices $i$ and $k$ refer to the embedding in the
higher-dimensional space. The GEMS approach is then valid if
the dimensionless parameter~$\lambda$,
\begin{equation}
\label{lambda}
\lambda = \frac{\sqrt{\Sigma^2}}{a^2}\,,
\end{equation}
is much smaller than one. If this condition is not satisfied, the GEMS
approach cannot provide information about the temperature.
%, though it
%may still be possible that a reliable measure of temperature
%exists. 

%--------------------------------------------------------------------
\subsection{Particles in AdS}
\subsubsection{Static particles}

To illustrate the surface gravity and GEMS approaches, let us as a
simple example first consider a static particle in an empty AdS space
(we will use AdS${}_5$ but generalisation to AdS in different
dimensions is straightforward). Staticity here can be defined with
respect to either the Poincar\'e or the global time. The orbits of a
static particle in these two coordinates are not equivalent (since the
two times are not the same).

The AdS metric in Poincar\'e and global coordinates, respectively, are
given by (we will use the mostly minus convention for the metric
signature throughout this paper):
\begin{equation}
\label{PoincareAds}
\begin{aligned}
{\rm d}s_5^2 &=  \frac{u^2}{R^2}({\rm d}t^2 - {\rm d}x_1^2 
-{\rm d}x_2^2 -{\rm d}x_3^2) - \frac{R^2}{u^2}{\rm d}u^2 \\[1ex]
&= R^2\left(  \cosh^2 \rho\ {\rm d}\tilde{t}^2 - {\rm d}\rho^2 -
\sinh^2 \rho\ {\rm d}\Omega_3 \right)\,.
\end{aligned}
\end{equation}
In global coordinates the surface gravity approach trivially gives
zero for a particle at rest, i.e.~following an orbit of the Killing
vector field $\tilde{\varsigma} = \partial_{\tilde{t}}$. The reason is that
this particle, even though it is not inertial (has non-zero proper acceleration),
 does not see any
horizon. We note that this is different with respect to flat space,
since in this case any observer with constant acceleration sees a
Rindler horizon and feels a temperature. 
%We will explicitly see below
%that in AdS space, there is a whole family of non-inertial
%observers, which experience a vanishing temperature.

In Poincar\'e coordinates, the situation is a bit different as the
particle sees a Poincar\'e horizon at $u=0$. An explicit computation for
this path yields that $k_H^2 = u^2/R^4$ which, using Tolman's law, yields
zero temperature.

It was shown in~\cite{Deser:1998xb} that the surface gravity
computation for the global coordinates is in agreement with the
higher-dimensional Unruh (i.e.~GEMS) approach.  The GEMS embeddings
for the zero temperature AdS space in Poincar\'e and global
coordinates are given in appendix~\ref{a:GEMS_AdS}. The
six-dimensional velocities and accelerations in the embedding space
are given by
\begin{equation}
\begin{aligned}
u &= \Big(-\frac{t}{R},0,0,0,-\frac{t}{R},1\Big)\,, \\[1ex]
a &= \Big(-u_0^{-1},0,0,0,-u_0^{-1},0\Big)\,,
\end{aligned}
\end{equation}
for a particle situated at the position $u= u_0 =\text{const}.$, $x_i
= \text{const}.$ in Poincar\'e coordinates.  Similarly,
\begin{equation}
\begin{aligned}
u &=  \Big( - \sin \tilde{t},0,0,0,0, \cos \tilde{t} \Big)\,, \\[1ex]
a &=  \Big(- \frac{\cos \tilde{t}}{R \cosh \rho_0}, 0,0,0,0, - \frac{\sin \tilde{t} }{R \cosh \rho_0 } \Big) \, , 
\end{aligned}
\end{equation}
for a particle situated at the position $\rho=\rho_0=\text{const}$,
$\theta, \phi_1, \phi_2=\text{const}.$ in global coordinates. While
the norm of the acceleration vector ($a^2 = - g_{ij} a^i a^j$) for the
static particle Poincar\'e AdS is zero, it is nonvanishing for the
stationary particle in global coordinates and equal to $a^2 = - 1/(R^2
\cosh \rho_0^2)$. Hence, while the Unruh relation~\eqref{accelT}
correctly yields the result that the temperature of a static observer
in Poincar\'e coordinates is zero, its naive application leads to a
non-zero and imaginary (!)  temperature in global coordinates.  What
happens is that for $a^2<0$, the trajectory of the particle in the
GEMS cannot be associated to a hyperbolic motion.  The detector-based
computation of~\cite{Deser:1998xb,Deser:1997ri} has shown that such an
observer would indeed measure zero temperature. One might think that
this happens because the detector sees no event horizon, but this
argument is at most heuristic, since e.g.~a detector which is
uniformly accelerated during a finite amount of time also does not see
any event horizon, but does measure excitations.  Note that because
AdS is globally non-hyperbolic, the measured temperature will
generically depend on the boundary conditions at infinity. We
emphasise that in the context of AdS/CFT, the appropriate boundary
conditions to use are the ``reflective'' ones (in the terminology
of~\cite{Avis:1977yn}), see e.g.~\cite{Nunez:2003eq}.

\subsubsection{Accelerated particles}
\label{s:planarAdSaccel}

Let us now turn to the more interesting case of accelerated particles,
that is to say, particles which move in empty AdS space in Poincar\'e
coordinates with a constant \mbox{acceleration}. More precisely, we
will consider particles which move such that the norm of the
\mbox{4-acceleration} in the directions parallel to the boundary of
AdS is constant.  Explicitly the path of the particle is:
\begin{equation}
 x_1^2 - t^2 = a_4^{-2} \, , \quad u =
  u_0 = \text{const}. \, , \quad x_2 = x_3 = 0 \,.
\end{equation}
The proper 5-acceleration for this particle is given by
\begin{eqnarray}
 a_5^2 = R^{-2} \left( 1 + \frac{a_4^2R^4}{
  u_0^2} \right) \,.
\end{eqnarray}
The time-like Killing vector which is associated with this trajectory
is~$\varsigma = \partial_\eta$, given in Rindler coordinates on the
four-dimensional slice (i.e.~$x_1 = \xi \cosh(\kappa\, \eta)$ and 
$t = \xi \sinh (\kappa\, \eta)$, such that an observer at constant
$\xi=\xi_0,\ x_2=x_3=0$ has $a_4=\xi_0^{-1}$). 
The accelerated particle observers see the Rindler horizon
at~$\xi=0$. Using the surface gravity computation we find
$k_H^2 = \kappa^2$.
Therefore, the temperature measured by the detector is
\begin{eqnarray}
T = \frac{a_4 R}{2\pi\, {u_0}} = \frac{1}{2 \pi } \sqrt{a_5^2 - R^{-2}}\,.
\label{TaccAds}
\end{eqnarray}
This is the same result as obtained in~\cite{Deser:1998xb} for an
accelerated particle in AdS space in global coordinates.

Turning now to the GEMS computation of the temperature, we first
determine the velocity and acceleration in the embedding space. The
path under consideration leads to
\begin{equation}
\begin{aligned}
u&=\Big(0,a_4 t,0,0,0,a_4\sqrt{a_4^{-2} + t^2} \Big)\,, \\
a&=\Big(0,\frac{a_4^2\,R}{ u_0}\sqrt{a_4^{-2} + t^2},0,0,0,
\frac{a_4^2\,t\,R}{ u_0}\Big) \, .
\end{aligned}
\end{equation}
The  modulus of the acceleration in the embedding 
6-dimensional space ($a_6^2=-g_{ij}a^ia^j$) is:
\begin{equation}
a_6 = \frac{a_4 R}{ u_0}\,.
\end{equation}
From $T=a_6/2\pi$ we then, consistently, recover the
result~\eqref{TaccAds}. 

\bigskip

%--------------------------------------------------------------------
\subsection{Strings in AdS}
\label{s:stringsAds}

Having discussed the effect of velocity and acceleration on the Unruh
temperature observed by particles in 
AdS, let us now turn to
strings. We will first consider Wilson loops, i.e.~string
configurations corresponding to non-dynamical (i.e.~infinite mass)
quark-antiquark pairs, and will then generalise these to finite-mass
meson configurations. The endpoints of a Wilson loop can be
accelerated either longitudinally, i.e.~along the direction of
separation of the endpoints (\S\ref{s:Wilson_long}), or in a direction
orthogonal to the separation (\S\ref{s:Wilson_ortho}). For the latter
case we compare accelerated loops at zero temperature with static
loops at finite temperature. We then consider more complicated
accelerated situations, namely finite quark-mass situations
(\S\ref{s:accelerated_wilson_loops}) and rotating mesons
(\S\ref{s:rotating_accelerated_mesons}).

\subsubsection{Longitudinally accelerated Wilson loops}
\label{s:Wilson_long}

Let us consider a Wilson loop whose endpoints are separated in
the~$x_1$ direction, and which accelerates in that same
direction.\footnote{In flat space, a string which extends in the
  direction in which it moves also experiences an acceleration bound,
  obtained by demanding that the endpoints do not lose causal
  contact. This situation is closely related to Bell's space-ship
  ``paradox''~\cite{Bell:1987hh}, and has been analysed in a large
  number of papers (see
  e.g.~\cite{Gasperini:1992jz,McGuigan:1994tg}). The
  difference with our setup here is that we have an extra,
  holographic, direction in which the string extends.} In the
following, it will be convenient to define a different radial variable
in the Poincare AdS metric (\ref{PoincareAds}), namely:
\begin{equation}
u=\frac{R^2}{z}\,.
\end{equation}
To describe such a string only the $t,x_1,z$
coordinates will play a role.  The relevant part of the AdS metric is
\begin{equation}
{\rm d}s^2 = \frac{R^2}{z^2}({\rm d}t^2 - {\rm d}x_1^2 - {\rm d}z^2)
+\dots
\end{equation}
To describe acceleration, it is convenient to make a change to Rindler coordinates 
\begin{equation}
x_1= \xi \cosh(\kappa \eta)\,,\qquad t= \xi \sinh(\kappa \eta)\,,
\end{equation}
such that a particle at constant $\xi$ is uniformly accelerated,
with 4-acceleration~$\xi = a^{-1}$. The metric is then
\begin{equation}
{\rm d}s^2 = \frac{R^2}{z^2}(\xi^2 \kappa^2 {\rm d}\eta^2 - {\rm
  d}\xi^2 - {\rm d}z^2)\,+\dots
\end{equation}
This metric has two horizons: the Poincar\'e horizon at $z\rightarrow
\infty$ and the Rindler horizon at $\xi=0$. We want to consider a
string which is static with respect to the $\eta$-time. The
accelerations of all points on the string worldsheet are thus constant
(but not equal to each other).

\begin{figure}[t]
\begin{center}
\includegraphics[width=.3\columnwidth]{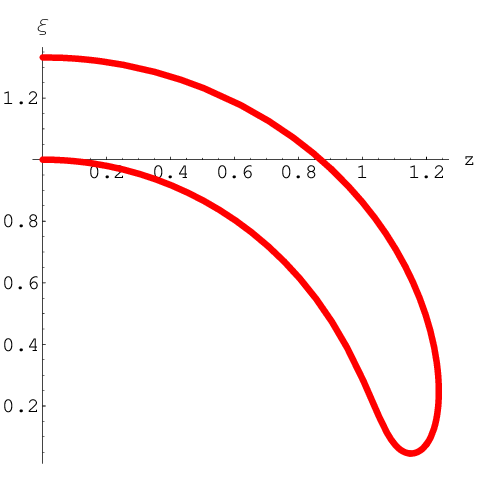}
\includegraphics[width=.3\columnwidth]{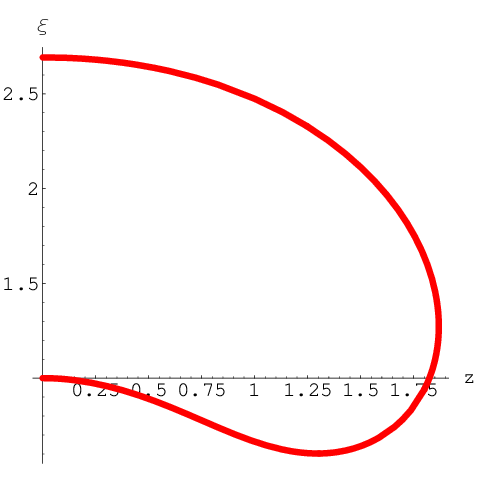}
\includegraphics[width=.3\columnwidth]{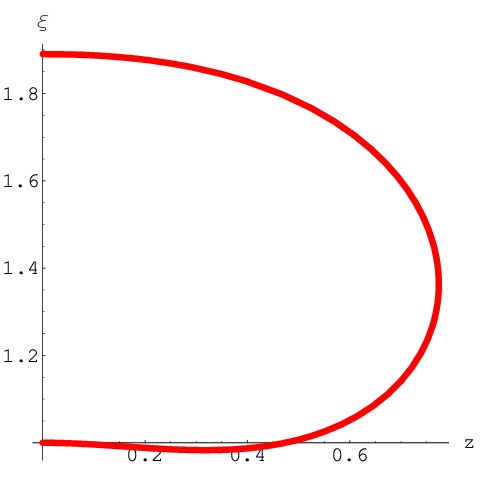}
\end{center}
\caption{Numerical solutions of a Wilson loop accelerated in the plane
  of the loop, with $a_L=1$. From left to right, $\xi_{3,L}=.01,.1,1$.}
\label{weirdfigure}
\end{figure}

Going to the static gauge $\tau=\eta$, $\sigma=z$,
$\xi = \xi(z)$, we write the Nambu-Goto 
action as:
\begin{equation}
S=-\frac{\kappa R^2}{2\pi \alpha'} \int\! {\rm d}z \, \frac{\xi}{z^2}\sqrt{1+\xi'^2}\,.
\label{acti}
\end{equation}
The equation of motion takes the form,
\begin{equation}
z (1 + \xi'^2) + \xi (2 \xi' + 2 \xi'^3  - z \xi'')  = 0 \, .
\label{zofxidiffeq}
\end{equation}
We want to solve this equation by imposing the boundary conditions that
the string endpoints are located at the boundary $z=0$ 
and move with accelerations $\xi_L(z=0) = a_L^{-1}$ and
$\xi_R(z=0) = a_R^{-1}$.

Expanding equation~\eqref{zofxidiffeq} around $z=0$ one gets a
two-parameter family of solutions describing the motion of each end of
the string,
\begin{equation}
\label{e:xiLR}
\xi_{L/R} = a_{L/R}^{-1} - \frac12 a_{L/R} 
z^2 + \xi_{3,L/R}\ a_{L/R}^2 z^3 - \frac18 a_{L/R}^3 z^4 + \dots\,,
\end{equation}
where $a_{L/R}^{-1}$ and $\xi_{3,L/R}$ are integration
constants. Regularity of the solution uniquely fixes $a_R$ and
$\xi_{3,R}$ in terms of the parameters on left-hand side, which means
that (as expected) the second order equation has two independent
parameters. In fact, because of the underlying rescaling 
symmetry, $a_R/a_L$ and $\xi_{3,R}$ are just functions of $\xi_{3,L}$.

Hence the nature of the solution only depends on the value of the
parameter $\xi_{3,L}$. When solving the differential
equation for $\xi_{3,L} >0$, one finds that the other string endpoint
reaches $z=0$ at a value of $\xi\equiv a_{R}^{-1} > a_{L}^{-1}$. 
Conversely, if $\xi_{3,L} < 0$, then $a_{R}^{-1} < a_{L}^{-1}$
($\xi_{3,L}$ and $\xi_{3,R}$ always have opposite sign).
Some
examples are depicted in figure~\ref{weirdfigure}.\footnote{The static
  gauge described above cannot be used to obtain the complete string
  shapes, as the solution is multi-valued. In practise, one thus has
  to change between the~$z=\sigma$ and~$\xi=\sigma$ gauges while
  solving the differential equations numerically.} By taking different
values of $\xi_{3,L}$ we can probe all the possible values of the ratio
between the acceleration of both endpoints.  

\begin{figure}[t]
\begin{center}
\includegraphics[width=.5\columnwidth]{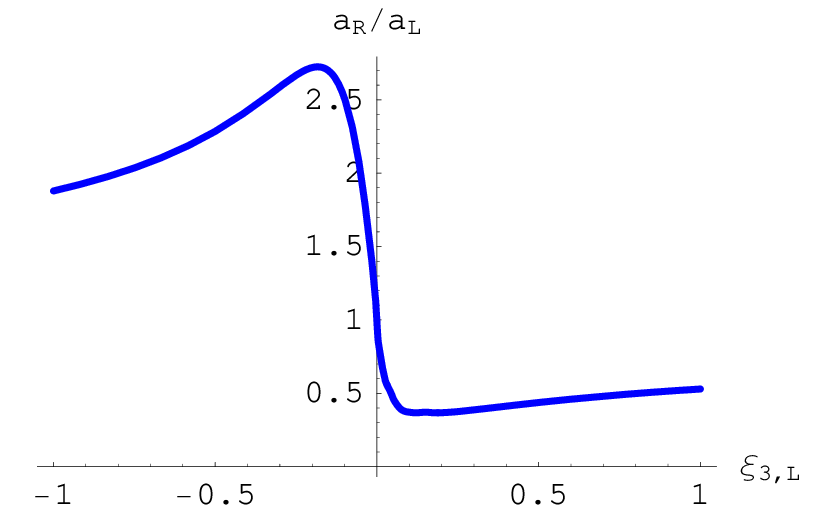}
\end{center}
\caption{The ratio between the acceleration of the string endpoints of
  a Wilson loop accelerated in the plane of the loop, as a function of
  the parameter $\xi_{3,L}$ which appears in~\protect\eqref{e:xiLR}. There
  is a minimum value of this ratio of approximately
  $\left(a_R/a_L\right)_{\text{min}}\approx 0.37$ and therefore a maximum of
  $\left(a_R/a_L\right)_{\text{max}}\approx 0.37^{-1}$. }
\label{aqoveraq}
\end{figure}

Curiously, there is a minimal (and, thus, a maximal) value of this
ratio, i.e.~a maximal separation between the string endpoints, see
figure~\ref{aqoveraq}. This maximal value is the analogue of the
screening length for orthogonally accelerated Wilson loops to be
discussed in~\S\ref{s:Wilson_ortho}.  Following the analogy
with~\S\ref{s:Wilson_ortho}, or with the well-known finite temperature
case~\mbox{\cite{Rey:1998bq,Brandhuber:1998bs}}, we expect that, to the right
of the minimum of the graph (and to the left of the maximum),
solutions should be perturbatively stable.  Between the two local
extrema, the string solutions should be unstable.\footnote{Perturbative
  stability of string configurations associated to Wilson loops was
  studied in~\cite{Avramis:2006nv}; see also~\cite{Friess:2006rk}.}  There
should also exist a point where the string becomes metastable towards
decay into two separate strings. We will not, however, pursue these
issues further.

\subsubsection{Accelerated single quark solution}
\label{s:single_quark}

By looking at figure~\ref{weirdfigure} we see that as~$\xi_{3,L}$
approaches~0, the endpoint accelerations approach each other,
$a_L\rightarrow a_R$, and the acceleration of the midpoint of the
string increases.  As the separation goes to zero ($\xi_{3,L}=0$), the
acceleration of the midpoint becomes infinite and we are effectively
left with a single trailing string, which asymptotically touches the
Rindler horizon. In other words, when $\xi_{3,L}=0$ the Wilson loop
``melts'', and dissociates into two (overlapping) isolated quarks.  In
this case it is actually possible to find a simple explicit solution
of (\ref{zofxidiffeq}):
\begin{equation}
\xi=\sqrt{a_q^{-2} - z^2}\, ,
\label{simplesol}
\end{equation}
where $a_q$ is an integration constant which is identified with the
acceleration of the isolated quark, because of the condition
$\xi|_{z=0} = a_q^{-1}$. One end of the string is attached to the AdS
boundary and the other one sits at the Rindler horizon.  The profile
(\ref{simplesol}) is, when compared to the finite temperature case of
\cite{Rey:1998bq,Brandhuber:1998bs}, the analogue of a straight string
hanging from the boundary to the AdS black hole
horizon.\footnote{After completion of this work we were informed that
  this solution has previously been found in a different
  context~\cite{Dominguez:2008vd}, and was further analysed
  in~\cite{Xiao:2008nr}, where the connection to the Unruh effect was
  pointed out (see~\cite{Chernicoff:2008sa} for a numerical analysis
  of generalisations of this solution to finite temperature). However,
  the global properties of their solution are different and lead to a
  horizon on the worldsheet. Our analysis gives an alternative
  interpretation of the exact solution as a limit of the numerical
  solutions discussed in the previous section.}

By inserting (\ref{simplesol}) into  (\ref{acti}), we can compute the
on-shell action of this string, which will be useful in
\S~\ref{s:Wilson_ortho}:
\begin{equation}
S_{\text{on-shell}}=
-\frac{\kappa R^2}{2\pi \alpha'\,a_q} \int_0^{a_q^{-1}}\! \,
 \frac{ {\rm d}z}{z^2}\,.
\label{actionshell}
\end{equation}
As expected,
this integral is divergent due to the infinite quark mass.

%--------------------------------------------
\subsubsection{Orthogonally accelerated Wilson loops: temperature vs.~acceleration}
\label{s:Wilson_ortho}

For particles, we know that temperature and linear acceleration are
interchangeable, by virtue of the Unruh effect. The situation is more
subtle for strings as different points on a string in general have
different acceleration. However, ignoring for the moment the details
of the string profiles, one may try to compare various physical
quantities, like for example screening lengths, in situations where
they are caused either by thermal effects or by acceleration.  In the
present section we will compute the possible profiles of a Wilson loop
accelerated in the $x_1$ direction, orthogonal to the quark separation
along $x_2$, and compare it to the results for a static Wilson loop in
a finite-temperature background~\cite{Rey:1998bq,Brandhuber:1998bs}.

\begin{figure}[t]
\begin{center}
\includegraphics[width=.46\columnwidth]{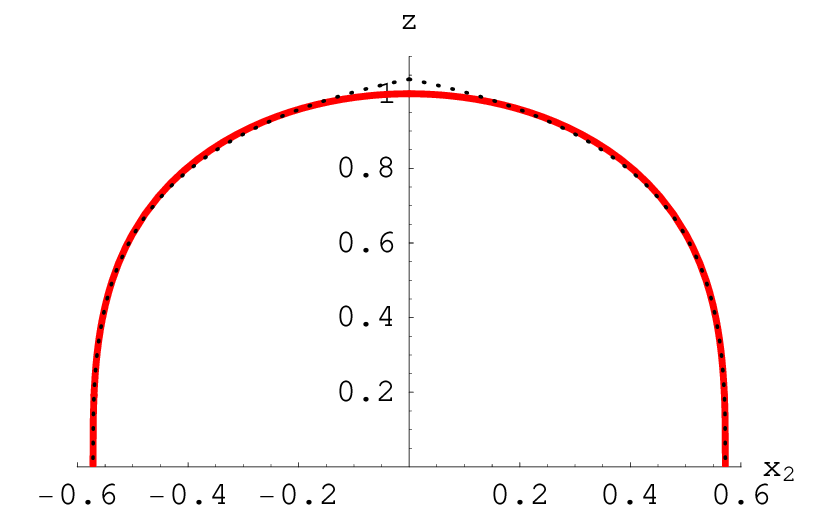}
\includegraphics[width=.46\columnwidth]{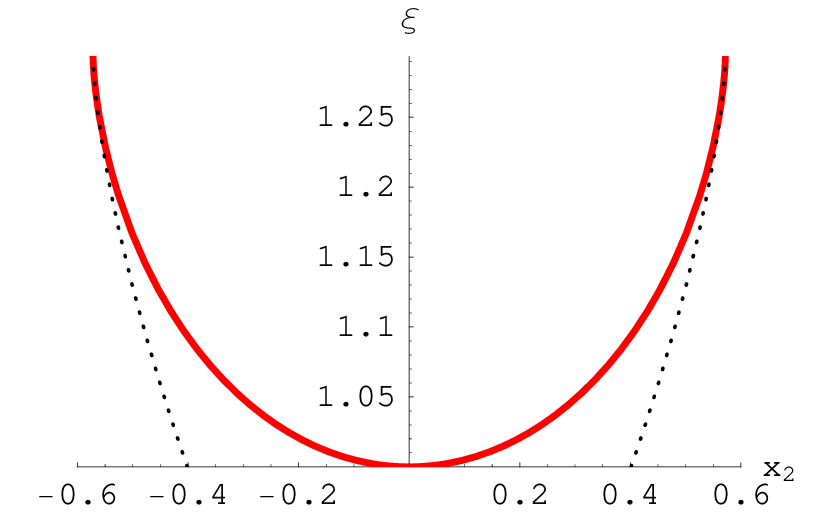}
\end{center}
\caption{Profile of a Wilson loop accelerated in a direction
  orthogonal to the plane of the loop,  with $z_0=\xi_0=1$. For this string,
  $L=1.14,\ a=0.77$.  The dotted line shows the asymptotic value of
  the functions near $x_2=L/2$ according to~\protect\eqref{boundaryexp}.}
\label{example1}
\end{figure}

We write the relevant part of the metric as
\begin{equation}
{\rm d}s^2 = \frac{R^2}{z^2}(\xi^2 \kappa^2 {\rm d}\eta^2 - {\rm
  d}\xi^2 - {\rm d}x_2^2- {\rm d}z^2) +\dots\,.
\end{equation}
We look for a solution static in~$\eta$. Using the
gauge~\mbox{$x_2=\sigma$} (so primes denote derivatives with respect
to~$x_2$), the Nambu-Goto action reads:
\begin{equation}
S=-\frac{\kappa R^2}{2\pi \alpha'} \int\! {\rm d}x_2 \, \frac{\xi}{z^2}\sqrt{1+\xi'^2+z'^2}\,.
\label{actiortho}
\end{equation}
(We have also performed our numerical analysis using the Polyakov
form, along the lines of~\cite{Herzog:2006gh}, which has the
disadvantage of an explicit constraint but the advantage that one can
easily enforce the world-sheet coordinates to be regular).
Considering a string symmetric with respect to its
turning point (which we choose to be at~\mbox{$x_2=0$}) such that
$z'|_{x_2=0}=\xi'|_{x_2=0}=0$, the equations of motion read:
\begin{equation}
\label{eqsxz}
\xi''=\frac{z_0^4 \xi}{\xi_0^2 z^4}\,\,,\qquad
z''=-\frac{2z_0^4 \xi^2}{\xi_0^2 z^5}\,\,,\qquad
\sqrt{1+\xi'^2+z'^2} = \frac{z_0^2 \xi}{\xi_0 z^2}\,\,.
\end{equation}
Here the last equation in (\ref{eqsxz}) is a consequence of the
existence of a conserved quantity, due to the fact that the Lagrangian
does not explicitly depend on~$x_2$. The two integration constants~$z_0$
and~$\xi_0$ correspond to the position of the midpoint of the string:
\begin{equation}
z|_{(x_2=0)}= z_0\,\,,\qquad\qquad \xi|_{(x_2=0)}= \xi_0\,.
\end{equation}
Equivalently these can be traded for~$L$ and~$a$, the
string length\footnote{Here and in the following, the term string length
should be understood as the separation between the endpoints in the direction
transverse to acceleration.}
and the acceleration of the string endpoints:
\begin{equation}
z|_{(x_2=L/2)}= z|_{(x_2=-L/2)}= 0\,\,,
\qquad\qquad \xi|_{(x_2=L/2)}= \xi|_{(x_2=-L/2)}= a^{-1}\,.
\end{equation}
It is interesting to see how equations~\eqref{eqsxz} behave near the
AdS boundary~$z=0$. At that point, $x_2$ reaches a finite value. 
By expanding the above equations we find
(we just focus on the $x_2=L/2$ endpoint, expansions around
$x_2=-L/2$ are trivially obtained from the ones below due to the symmetry
$x_2 \to -x_2$):
\begin{equation}
\xi=a^{-1} - \frac{3^\frac23 z_0^\frac43 a^\frac13}{2\xi_0^\frac23}
(\frac{L}{2}-x_2)^\frac23+\dots\,\,,\qquad
z= \frac{3^\frac13 z_0^\frac23}{a^\frac13\xi_0^\frac13}
(\frac{L}{2}-x_2)^\frac13-\frac{3a\,z_0^2}{5\xi_0}(\frac{L}{2}-x_2)+
\dots\,\,.
\label{boundaryexp}
\end{equation}
Given any pair $\xi_0,z_0$ we can numerically integrate the above
equations and determine $L, a$.  Figure~\ref{example1} depicts an
example.

\begin{figure}[t]
\begin{center}
\includegraphics[width=.5\columnwidth]{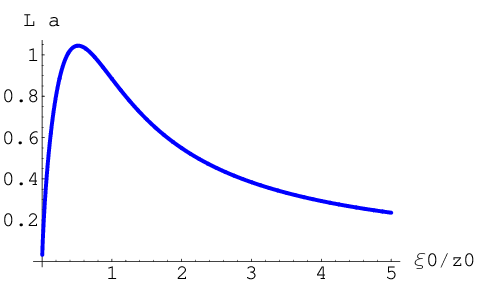}
\end{center}
\caption{$L\,a$ as a function of $\xi_0/z_0$ for an orthogonally
  accelerated Wilson loop.}
\label{Laplot}
\end{figure}

Due to the underlying conformal symmetry,
 a simultaneous rescaling $\xi_0 \to \lambda \xi_0 $ and
$z_0\to \lambda z_0$ will just rescale the whole solution,
yielding~\mbox{$L \to \lambda L ,a\to \lambda^{-1} a$}. Thus, for
fixed $\xi_0/z_0$ there is a fixed value of~$L\, a$.  See
figure~\ref{Laplot}. From this graph we conclude that, for a fixed
acceleration, the maximum quark-antiquark separation for which a 
connected string solution of the Nambu-Goto action exists is:
\begin{equation}
\label{LmaxAdS}
L_{\text{max}} = \frac{1.045}{a}\,\,,\qquad\qquad
\frac{\xi_0}{z_0}\Big|_{L_{\text{max}}} = 0.516\,\,.
\end{equation}
In the language of the dual gauge theory, the colour force is screened
for larger separation.  We can also define a critical length at which
the accelerated connected string becomes energetically disfavoured
and, thus, metastable, with respect to two separate accelerated
strings, see \S~\ref{s:single_quark}.  Even if the energy integrals
are divergent, we can define a renormalised energy as the energy of
the connected string minus twice the energy of a single quark string,
\begin{equation}
E_{\text{ren}}= \frac{\kappa R^2}{2\pi \alpha'}
\left[ 2\int_0^{z_0}\! {\rm d}z \, \frac{\xi^2z_0^2}{z^4\xi_0\,z'}
-2 a^{-1}\int_0^{a^{-1}}\! \frac{{\rm d}z}{z^2} \right]\,,
\end{equation}
where we have inserted (\ref{actionshell}). 
Using the expansions (\ref{boundaryexp}) one can check that the divergences
around $z=0$ cancel out. The critical length is defined as the one for
which $E_{\text{ren}}=0$. A numerical computation produces the following
approximate results:
\begin{equation}
\label{LcritAdS}
L_{\text{crit}} = \frac{0.90}{a}\,\,,\qquad\qquad
\frac{\xi_0}{z_0}\Big|_{L_{\text{crit}}} = 0.96\,\,.
\end{equation}

We now proceed to compare these results to the case of static strings
in the AdS black hole \cite{Rey:1998bq,Brandhuber:1998bs}.  For this
purpose, it is useful to notice that, using~\eqref{TaccAds}, the ratio
$\xi_0/z_0$ can be related to the Unruh temperature felt by the
midpoint of the string, namely:
\begin{equation}
2 \pi T_{\text{mid}} = \frac{z_0}{R\,\xi_0}\,.
\end{equation}
Note that this temperature has a clear-cut meaning on the string
theory side, but does not necessarily translate in a straightforward
way to the gauge theory side, in contrast to the endpoint temperature.
We also want to compare our result to the case of a string accelerated
in flat space~\cite{Frolov:1990ct,Berenstein:2007tj}.  We remind the
reader of the relation, in that case, between the string length, the
acceleration of the string endpoints (which we denote by~$a$) and the
string midpoint acceleration,
\begin{equation}
a=\frac{a_{\text{\text{mid}}}}{\cosh\left(
\frac{a_{\text{\text{mid}}}L}{2}\right)}\,\,,
\label{flatspacea}
\end{equation}
from which one finds the maximal length for fixed
acceleration $L_{\text{max}}=1.325/a$.

\begin{figure}[t]
\begin{center}
\includegraphics[width=.5\columnwidth]{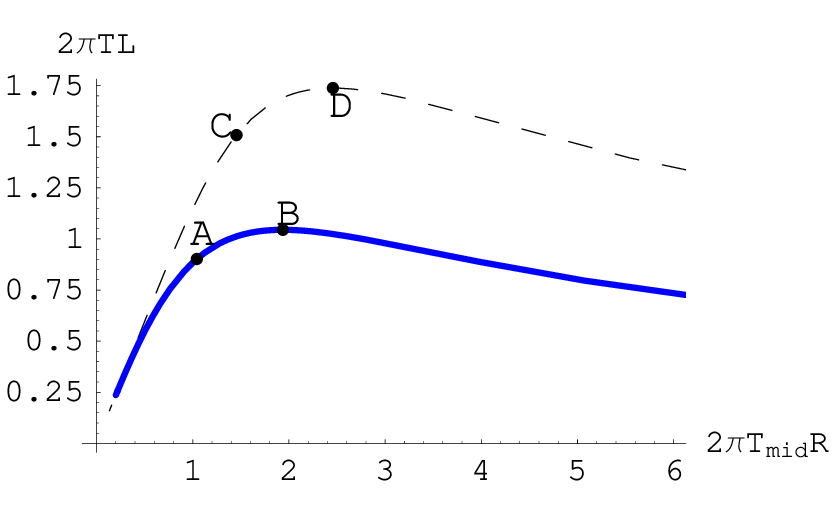}
\end{center}
\vspace{-2ex}
\caption{Comparison between the finite acceleration (solid line) and
finite temperature (dashed line) cases. 
Notice that the temperature plotted in the vertical axis is a temperature
in the dual field theory while the one in the horizontal axis is the temperature
felt by the string midpoint in the bulk theory.
The string solutions corresponding to points on the right of
B, D are perturbatively unstable; between A-B and C-D are metastable
and on the left of A, C, are stable.}
\label{figcompare}
\end{figure}

The comparison between the finite acceleration and finite temperature
cases is summarised in figure \ref{figcompare} and table \ref{tabcompare},
where we have also included the flat space result.
We see that acceleration dissociates extended objects more
efficiently than its associated Unruh temperature.

\begin{table}
\vspace{2ex}
\begin{center}
\begin{tabular}[!ht]{|c|c|c|}
\hline
\ & $2\pi T L_{\text{max}}$ & $2\pi T L_{\text{crit}}$ \\
\hline
accelerated string in AdS & 1.045 & 0.90 \\
\hline
static string in AdS black hole & 1.738 & 1.508 \\
\hline
accelerated string in flat space & 1.325 & 1.055 \\
\hline
\end{tabular}
\end{center}
\caption{Comparison of some quantities between the different setups.
  For the acceleration cases, the temperatures should be understood as
  Unruh temperatures, $T=a/(2\pi)$, where~$a$ is the acceleration of
  the endpoints. The entries on the first two lines of the table
  relate to the points B, A, D and C in
  figure~\protect\ref{figcompare}.  }
\label{tabcompare}
\end{table}
Notice that for this comparison, we need to relate the temperature of
the heat bath in the formula~\eqref{accelT} with some kind of
effective temperature felt by the accelerated string. This is in
principle tricky, as the string is an extended object, and different
points on the string have different accelerations, and in principle
feel different local Unruh temperatures. For mesons it seems natural
to consider the Unruh temperature of the quarks as a relevant
temperature, as this is the parameter that should be easiest to
control experimentally, given that quarks are charged.

%%%%%%%%%%%%%%%%%%%%%%%%%%%%%

\subsubsection{Orthogonally accelerated mesons}
\label{s:accelerated_wilson_loops}

\begin{figure}[t]
\begin{center}
\includegraphics[width=.46\columnwidth]{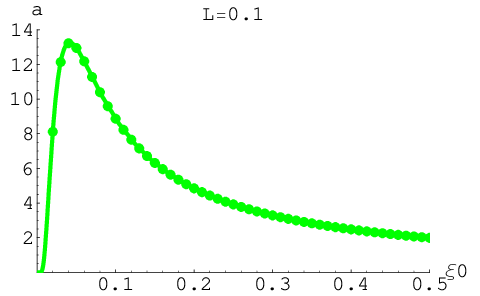}
\includegraphics[width=.46\columnwidth]{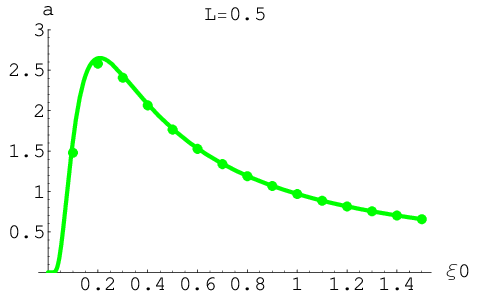}\\[2ex]
\includegraphics[width=.46\columnwidth]{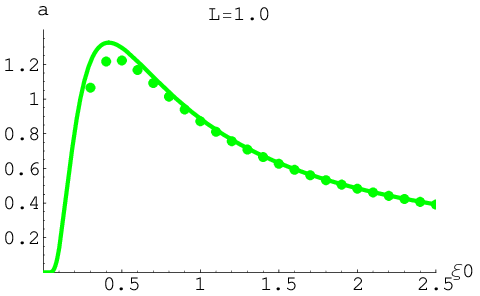}
\includegraphics[width=.46\columnwidth]{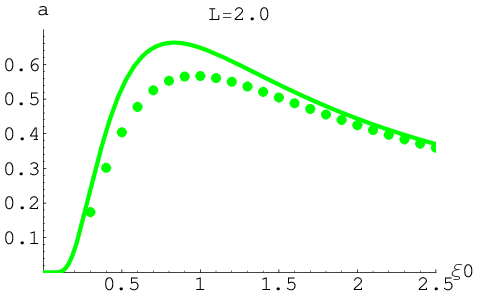}
\end{center}
\vspace{-2ex}
\caption{The endpoint acceleration for a Wilson loop suspended from
  the~$z_m=1$ brane in AdS, as a function of the inverse midpoint
  acceleration~$\xi_0=  a_{\text{mid}}^{-1}$, for various values of the endpoint
  separation~$L$. The continuous curves depict the flat-space
  relation~\protect\eqref{flatspacea}.\label{f:a_vs_xi0}}
\end{figure}

We now want to consider acceleration of finite-energy string
configurations, corresponding to mesonic states with finite
constituent quark masses. The simplest way of obtaining information
about mesons with dynamical quarks is to consider the regularised
version of the Wilson loop. That is, we will solve the same equations of
motion for a U-shaped string of \S~\ref{s:Wilson_ortho}, 
but with endpoints fixed at some probe brane $z=z_m$.
Then, we compute the length and acceleration of the mesonic string as
\begin{equation}
z|_{(x_2=L/2)}= z|_{(x_2=-L/2)}= z_m\,\,,
\qquad\qquad \xi|_{(x_2=L/2)}= \xi|_{(x_2=-L/2)}= a^{-1}\,.
\end{equation}
The relation we are after is the behaviour of the maximal
acceleration~$a_{\text{max}}$ of these endpoints as a function of
their fixed separation~$L$. We will assume that appropriate forces
have been introduced to keep this separation constant at all times,
and we will thus discard any boundary terms in the equations of
motion.  This situation is as close as one can get to the acceleration
bound found in~\cite{Peeters:2007ti} when one considers non-rotating
configurations (which have their size stabilised dynamically).

\begin{figure}[t]
\begin{center}
\includegraphics[width=.45\columnwidth]{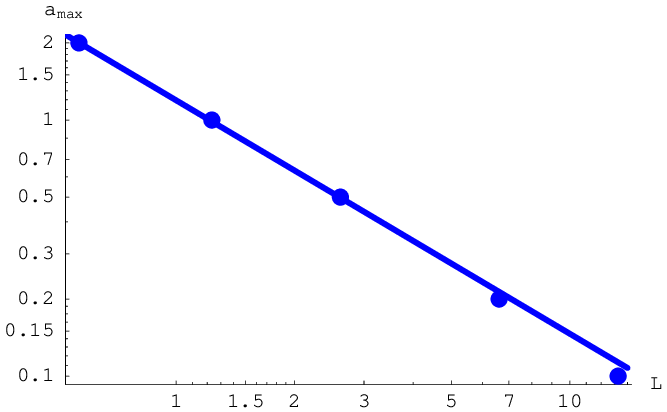}
\end{center}
\caption{The relation between the endpoint
  acceleration~$a_{\text{max}}$ and the length~$L$ of the projection
  of the Wilson loop on the~$z=1$ brane (log-log plot).  The fitted
  straight line is \mbox{$a_{\text{max}} = 1.20/L^{0.91}$}.\label{f:amax_vs_L_loglog}}
\end{figure}

To find the relation~$a_{\text{max}}(L)$, we have determined the
endpoint acceleration for various separations as a function of the
midpoint acceleration. The result is displayed in
figure~\ref{f:a_vs_xi0} (where we have chosen the D7-brane to be
located at~$z_m=1$ for convenience). The points in these figures were
obtained by finding~$z_0$ such that the endpoint separation is equal
to the fixed value of~$L$ given in each plot, and then registering the
corresponding~$a$.

By analysing the position of the maximum in~$a$ as a function of~$L$,
one then finds figure~\ref{f:amax_vs_L_loglog}. This result is to be
compared to the analysis of~\cite{Frolov:1990ct,Berenstein:2007tj},
see~\eqref{flatspacea}.  Our results for the AdS setup are
qualitatively similar to this flat space result.

Alternatively, one can consider a similar U-shaped string
configuration where the separation between the endpoints is determined
dynamically, and stabilised by rotation of the string. To this
technically more involved configuration we turn now.

\subsubsection{Rotating accelerated mesons}
\label{s:rotating_accelerated_mesons}

Real world mesons have a non-vanishing angular momentum, and hence it
is also interesting to analyse the behaviour of rotating string
configurations in AdS. 
The relevant piece of the metric is
\begin{equation}
{\rm d}s^2= \frac{R^2}{z^2}\left(\kappa^2\xi^2 {\rm d}\eta^2 - {\rm d}\xi^2 -
{\rm d}\rho^2 - \rho^2 {\rm d}\phi^2
-{\rm d}z^2 \right)\,+\dots
\label{rotating_metric}
\end{equation}
For the rotating string, the natural ansatz is,
in terms of the worldsheet coordinates $\tau, \sigma$,
\begin{equation}
\eta = \tau\,\,,\quad
\phi =\kappa\, \omega\, \tau \,\,,\quad
\xi = \xi (\sigma)\,\,,\quad
\rho = \rho (\sigma)\,\,,\quad
z = z (\sigma) \,\,.
\end{equation}
Notice that we have already fixed part of the worldsheet
reparametrisation invariance. This ansatz immediately solves the
equations and constraints for $\eta$ and $\phi$.  We are left with a
reduced Nambu-Goto action,
\begin{equation}
S= - \frac{R^2\,\kappa}{2\pi \alpha'}
\int\!{\rm d}\tau {\rm d}\sigma\, \frac{1}{z^2}\sqrt{\xi^2 - \rho^2 \omega^2}
\sqrt{\xi'^2 + \rho'^2 + z'^2}\,,
\label{reducedNG}
\end{equation}
where a prime denotes a derivative with respect to~$\sigma$.
The equations of motion can be readily obtained from this action.
An important quantity is the angular momentum,
\begin{equation}
J = \frac{R^2\,\kappa}{2\pi \alpha'}
\int\!{\rm d}\sigma\, \frac{1}{z^2} \rho^2 \omega
\frac{\sqrt{\xi'^2 + \rho'^2 + z'^2}}{\sqrt{\xi^2 - \rho^2 \omega^2}}\,.
\end{equation}
Note that all these expression reduce to those
of~\cite{Peeters:2007ti} if we would insert $z=\text{const}.$ and
identify $T_s = (R^2\,\kappa)/(2\pi \alpha' z^2)$, although this is of
course not consistent in the present setting.

As in~\S~\ref{s:accelerated_wilson_loops}, we will consider strings
hanging from a brane at $z=z_m$.  Any solution of the reduced
action~\eqref{reducedNG} also solves the full Nambu-Goto
action. However, not all solutions are physically consistent since the
open string action has to be supplemented with a boundary
condition. This was first discussed for a similar set-up
in~\cite{Kruczenski:2003be} where it was shown that the string had to
hit the brane orthogonally. The argument is similar in the present
case and leads to:
\begin{equation}
\partial_z \rho|_{z=z_m} = 0 \,\,,
\label{constort}
\end{equation}
which, in turn, implies $\partial_\xi \rho|_{z=z_m} = 0$.
Thus, for fixed $z_m$, the family of solutions is therefore defined
by three parameters $\omega$, $\xi_0\equiv \xi|_{\rho=0}$ and
$z_0\equiv z|_{\rho=0}$. These three parameters are subject to 
one constraint (\ref{constort}), so we are left with a two-parameter
family. The two parameters can be identified with the physical 
quantities $J$ and $a$.
\begin{figure}[t]
\begin{center}
\includegraphics[width=.45\columnwidth]{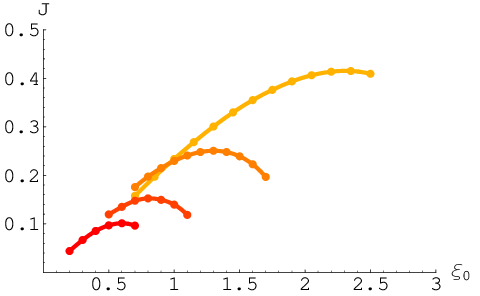}\quad
\includegraphics[width=.45\columnwidth]{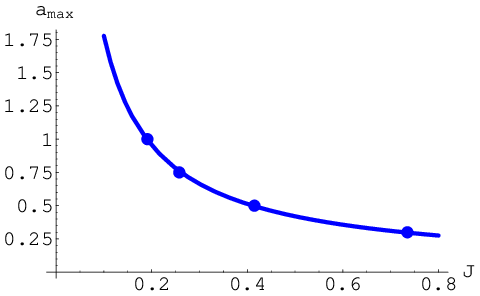}
\end{center}
\caption{Left: the relation between the angular momentum~$J$ and the
  inverse acceleration~$\xi_0$ of the midpoint of the string, for
  values of the endpoint acceleration~$a$ in the set~$\{ 0.3, 0.5,
  0.75, 1.0\}$ (the highest curve has the smallest value
  of~$a$). Right: the relation between the endpoint
  acceleration~$a_{\text{max}}$ and the value of the the angular
  momentum~$J_{\text{max}}$ for the maximally accelerated rotating
  string configuration at a given angular momentum. The fitted curve
  is~$a_{\text{max}} = 0.226/J^{0.895}$.
For the figures, we have set $z_m=1$ and
the angular momentum $J$ is measured in units of $\frac{R^2\,\kappa}
{2\pi\alpha'}$.
\label{f:J_vs_xi0}}
\end{figure}

We have numerically analysed families of configurations with different
values of the midpoint and endpoint acceleration. These curves exhibit
a maximum in~$J$, see figure~\ref{f:J_vs_xi0}a. For any given~$a$
there is thus a maximal~$J$, and the relation is plotted in
figure~\ref{f:J_vs_xi0}b.  Our main result is that there is an upper
bound on the acceleration (beyond which no mesons exist) which scales
with the angular momentum as
\begin{equation}
a_{\text{max}} \sim 1/J^\alpha\,,
\end{equation}
where~$\alpha$ is a positive number. This exponent~$\alpha$
equals~$1/2$ in flat space and grows to about~$0.89$ in the AdS
background which we considered here, at least in the region of 
$J$ displayed in figure \ref{f:J_vs_xi0}.

This concludes our analysis of accelerated strings in the
zero-temperature AdS background. We will now turn to the analysis of
strings in the finite-temperature AdS black hole background.

%====================================================================
\section{Velocity and acceleration effects in finite temperature backgrounds}
\label{s:va_finite_T}

Having discussed the Unruh effect for various observers in curved
space with vanishing Hawking temperature, we now want to consider
cases in which the background has a non-zero temperature. The problem
of an observer moving with constant velocity through a heat bath has
received considerable attention in the literature (in part because it
is relevant for the interpretation of measurements of the cosmic
microwave background), and we will here add some more ingredients to
this debate. Briefly, such an observer sees an angle-dependent
spectrum, which is thermal for each fixed angle, but with a
temperature which depends on the angle~$\theta$ between the direction
of motion and the direction of
observation~\cite{Bracewell:1968a,Peebles:1968zz,Henry:1969im}, see
also~\cite{Aldrovandi:1992ba},
\begin{equation}
T_{\text{obs}}(\theta) = \frac{T\sqrt{1-v^2}}{1- v \cos\theta}\,.
\end{equation}
As explained in~\cite{Costa:1995yv,Landsberg:1996ac}, there are thus
various ways to define an ``effective'' temperature (keeping in mind,
of course, that the spectrum really is not thermal to begin with). The
resulting power of~$\gamma=1/\sqrt{1-v^2}$ will depend on the details
of this averaging procedure~\cite{Costa:1995yv}. We will see in the
present section that the effective temperature observed by a moving
string, in the sense of the temperature relevant for its dissociation,
is in fact obtained simply from a surface gravity  computation.

%--------------------------------------------------------------------
\subsection{Constant velocity particles in the AdS-BH background}
\label{s:particle_velo_AdSBH}

As we have emphasised in the introduction, the temperature measured by
observers in finite temperature backgrounds depends in a more
complicated way on acceleration and velocity than in zero temperature
backgrounds. We will be interested in the case of particles and
strings moving parallel to a planar AdS black hole horizon.

The simplest case one can analyse is one in which a particle-like
observer moves with constant velocity at fixed distance~$u$ away from
a flat black hole in AdS (or in other words, on a D-brane suspended at
fixed distance from the horizon). The relevant metric is
\begin{equation}
\label{PoincareAdsBH}
{\rm d}s^2 = \Big( \frac{u^2}{R^2} h(u) {\rm d}t^2 - \frac{u^2}{R^2}\big(
{\rm d}x_1^2+{\rm d}x_2^2 + {\rm d}x_3^2\big) - \frac{R^2}{u^2 h(u)}{\rm d}u^2\Big)\,,
\end{equation}
with~$h(u) = 1 - (u_h/u)^4$, where $u_h$ is related to the
temperature of the dual field theory as $T=u_h/(\pi R^2)$. 
We will consider the orbit for which~$x_1 = v t$. The relation between
coordinate time and proper time is a function of~$u$,
\begin{equation}
\frac{{\rm d}\tau}{{\rm d}t} = \frac{u}{R} \sqrt{1 -
  \left(\frac{u_h}{u}\right)^4 - v^2}\,.
\end{equation}
The velocity is thus necessarily bounded by the local velocity of
light, \mbox{$v < \sqrt{1-(u_h/u)^4}$}. The Killing vector relevant
for the computation of the surface gravity is
\begin{equation}
\varsigma = \frac{\partial}{\partial t} - v \frac{\partial}{\partial x_1}\,.
\end{equation}
Evaluating the surface gravity expression~\eqref{kH} at the particle
horizon~$u = u_h (1-v^2)^{-1/4}$ we find 
\begin{equation}
k_H = 2 \frac{u_h}{R^2} (1-v^2)^{1/4}\,.
\end{equation}
For the observed temperature we use the Tolman law, which in this case
yields\footnote{The same temperature can be obtained from a different
  perspective, namely by considering coordinates in which the particle
  is static but feels a hot wind from the plasma moving at
  velocity~$v$. In order to check this, one can first boost the
  metric~\eqref{PoincareAdsBH} in the $x_1$ direction by changing
  ${\rm d}t \to (1-v^2)^{\frac{1}{2}}({\rm d}t+v {\rm d}x_1)$, ${\rm
    d}x_1 \to (1-v^2)^{\frac{1}{2}}({\rm d}x_1 +v {\rm d}t)$, which gives
\begin{displaymath}
 {\rm d}s^2 = \frac{u^2}{R^2}\left[\left(1-\frac{u_h^4}{u^4(1-v^2)}\right)
  {\rm d}t^2 
 - 2\frac{u_h^4 v}{u^4 (1-v^2)}{\rm d}t\,{\rm d}x_1 - 
\left(1+\frac{v^2 u_h^4}{u^4(1-v^2)}\right){\rm d}x_1^2
- 
{\rm d}x_2^2 - {\rm d}x_3^2\right] - \frac{R^2}{u^2 h(u)}{\rm d}u^2\,.
\end{displaymath}
Computing \eqref{kH} for this metric with $\varsigma = \partial_t$, one
finds $k_H=2u_h R^{-2} (1-v^2)^{-\frac{1}{4}}$. Upon insertion 
in \eqref{TL} this yields~\eqref{Tforconstantv}, as expected.}:
\begin{equation}
T = \frac{u_h (1-v^2)^{1/4}}{\pi R u \sqrt{\displaystyle 1 -
    \left(\frac{u_h}{u}\right)^4 - v^2}}\,.
\label{Tforconstantv}
\end{equation}
We should note several features of this formula. Firstly, we see that
in the $v\rightarrow 0$ limit we recover the known
result~\cite{Deser:1997ri,Brynjolfsson:2008uc} that a static observer
at asymptotic infinity of the AdS-Schwarzschild black hole measures a
vanishing temperature. This is in contrast with an asymptotic static
observer in a flat space black hole spacetime, which measures a
non-vanishing Hawking temperature.

Secondly, the behaviour of this temperature for large distance away
from the horizon is remarkable. In the limit~$u \gg u_h$ it scales
with the velocity as
\begin{equation}
\label{e:Tvflat}
T_{\text{obs}} = T_0 (1-v^2)^{-1/4},
\end{equation}
where~$T_0$ is the local temperature for a static observer at
position~$u$. This result is to be compared with the behaviour of the
screening length of the colour force, found in the string computations
of~\cite{Peeters:2006iu,Liu:2006nn,Chernicoff:2006hi}, and approximately given by:
\begin{equation}
\label{e:SL}
L_s = L_0  (1-v^2)^{1/4} / T_H \,.
\end{equation}
If we use~\eqref{e:Tvflat} at the centre-of-mass position of the
string, we have~$T_H \sim T_0$ and the relation~\eqref{e:SL} simply
states that the screening length is inversely proportional to the
observed temperature, \mbox{$L_s \sim 1/T_{\text{obs}}$}.
In~\cite{Caceres:2006ta}, this result was generalised to other
backgrounds different from $AdS_5$, the main difference being the
exponent of the $(1-v^2)$ factor. We have checked in a variety of
cases that the method presented in this section reproduces the results
of~\cite{Caceres:2006ta}. We want to emphasize that computations of
this section do not involve any stringy action.\footnote{The
  \emph{relevance} of the temperature scale~$T_{\text{obs}}$ has of
  course been pointed out many times before (see in this context
  also~\cite{Dominguez:2008vd} where the connection to transverse
  momentum broadening is discussed). Our computation shows \emph{why}
  this scale appears naturally.}

We can also try to compute the observed temperature by using an
embedding in a globally Lorentzian space-time. In order to construct
this embedding we need to suitably modify the embedding of the
spherical AdS black hole given in~\cite{Brynjolfsson:2008uc}, see
appendix~\ref{a:GEMS_AdS_BH}.  In this constant velocity case,
however, the jerk of the associated trajectory is not small and there
is no immediate interpretation from the GEMS. This corresponds to the
fact, already alluded to at the beginning of this section, that the
observed spectrum is only approximately thermal. It could prove
interesting to tackle this problems with methods similar
to~\cite{Chen:2004qw}, but we will not attempt that here.

%--------------------------------------------------------------------
\subsection{Accelerated particles in the planar AdS black hole}
\label{s:particle_accel_AdSBH}

For an accelerated string in an AdS black hole background, the
equations of motion are much more complicated than for the constant
velocity case discussed in the previous subsection. We will comment on
these string configurations in the next subsection. Before we do so,
let us first again try to get an intuition for the physics by
analysing the much simpler case of a point particle.

The point particle we will analyse here is one which is accelerating
parallel to the flat horizon of a planar AdS black hole. We want to
consider a particle with a constant 4-acceleration. This is
similar to the computation of~\S\ref{s:planarAdSaccel}, but now
for the black hole case. The particle follows the path
\begin{equation} 
x_1^2 - h(u )t^2  = a_4^{-2} \, , \qquad  \qquad  
u,x_2,x_3=\text{const}.\,. 
\end{equation}
Note that in order to ensure that the particle has a constant
4-acceleration $a_4$ we have to include the factor 
$h(u)\equiv 1 - (u_h/u)^4$, which
accounts for the fact that there is a non-trivial red-shift between two
different values of the radial coordinate.

We will attempt to compute the temperature observed by this
particle using the GEMS approach. 
 By using the
GEMS embedding of the planar AdS black hole~\eqref{PoincareBHGEMS2} we
compute the norm of the higher-dimensional GEMS acceleration,
\begin{equation}
a_8^2 = \frac{u^2}{R^2(u^4-u_h^4)}(4u_h^2 + a_4^2 \,R^4+ 8a_4^2 u_h^2 t^2
+ 4 a_4^4 u_h^2 t^4) \, .
\label{a8eq}
\end{equation}
We see that, unlike in all previous cases, this norm has a
\emph{polynomial dependence on time}.  The jerk modulus is non-zero
and also grows in time,
\begin{equation}
|\Sigma|=\frac{2u^2u_h a_4 \,t}{R^2 (u^4 - u_h^4)}
\sqrt{(1 + a_4^2 t^2)((a_4^2(9R^4 + 8 u_h^2 t^2) +  
4 u_h^2(1 + a_4^4 t^4))}\,.
\end{equation}
We can now compute the parameter $\lambda$~\eqref{lambda} which
measures the applicability of the GEMS approach,
\begin{equation}
\lambda^2 = \left(\frac{|\Sigma|}{
a_8^2}\right)^2 = 4  a_4^2 t^2 u_h^2
\frac{(1 + a_4^2 t^2)((a_4^2(9R^4 + 8 u_h^2 t^2) +  
4 u_h^2(1 + a_4^4 t^4))}{(4 u_h^2 + a_4^2\,R^4 + 8 a_4^2 u_h^2 t^2
+ 4 a_4^4 u_h^2 t^4)^2}  \, .
\end{equation}
We see that $\lambda$ is independent of the radial position of the
particle, but depends on time. Actually, we see that for $t \rightarrow \infty$ $\lambda
\rightarrow 1$, i.e.~it is not possible to choose parameters in such a
way that $\lambda$ remains small for all times.  

Nevertheless, notice that the jerk modulus and, accordingly,
$\lambda$, vanish at $t=0$ (actually the whole jerk vector vanishes at
this point, not just its modulus).  Thus, we can
consistently define an effective temperature that an accelerated
detector would observe at $t=0$, the instant at which the velocity
vanishes although the acceleration is finite\footnote{This is
  completely analogous to the discussion
  in~\cite{Brynjolfsson:2008uc}, where a temperature for a freely
  falling observer in a black hole background could only be defined at
  the instant when the observer is at rest. The notion of instantaneous
temperature is not well-defined and, in order for   (\ref{T2T2T2}) to coincide
with a temperature measured by our hypothetical detector, the typical time of
the microscopic processes involved in the detector measurements should be much
smaller than the critical time at which (\ref{T2T2T2}) loses its validity. In 
general, one should think of (\ref{T2T2T2}) as an effective approximate expression
for the amount of heating felt by the accelerated particle.}. Inserting $t=0$ in
(\ref{a8eq}) we reach the expression:
\begin{equation}
T_{\text{obs}}^2 = T_{\text{bh,static}}^2 + T_{\text{accel}}^2\,\,,\qquad\quad
(t=0)
\label{T2T2T2}
\end{equation}
where we have defined $T_{\text{bh,static}}$ as the temperature that
a static observer at the same constant $u$ would observe and
$T_{\text{accel}}$ as the temperature coming just from the observer acceleration
in the 4-d constant $u$ slice of the geometry
(the Unruh temperature for the same trajectory if in the metric 
(\ref{PoincareAdsBH}) one discards the d$u^2$ term).

Hence, we conclude that an accelerated particle in the background of a
planar AdS black hole, at the initial stage of the acceleration, is in
an approximate state of thermal equilibrium, with a local
temperature that increases in time. After a critical time~$t_c$, the
GEMS approach can no longer be used, and we cannot say what is the
final destiny of such a particle. The
  critical time is a complicated function of $a$ and $u_h$ which is a
  solution of a particular cubic equation $\lambda^2=\epsilon\ll 1$.

One may wonder if the surface gravity approach could be used more
successfully here. Unfortunately, the problem which one faces in this
case is that unlike in all previous cases, the path which the particle
follows is not generated by a Killing vector field, which prevents one
from using this method.

%--------------------------------------------------------------------
\subsection{Particles versus strings}

Finding even a numerical solution of a string with accelerated
endpoints in the background of an AdS black hole is a non-trivial
problem. The reason is that a generic string configuration with such
boundary conditions will be space-time dependent, and it is a priori
not clear that it is possible to find a \emph{globally} well-defined
time variable in which the string motion would be static. In contrast
to~\S\ref{s:accelerated_wilson_loops}, where we used Rindler
like coordinates to reduce the system of partial differential
equations to an ordinary differential equation, this is now no longer
possible.  In addition it is also not clear which one of the many
possible time-dependent solutions satisfying the boundary conditions
of accelerated endpoints is the most relevant one (e.g.~which one has
the lowest energy, and corresponds to an unexcited accelerated meson).

However, despite all above mentioned problems and subtleties, it seems
reasonable to expect that the qualitative behaviour of the accelerated
meson can be extracted by considering an accelerated point particle
instead of string, as a first approximation. To see what is the
difference of the full (string) solution versus the particle
approximation, let us consider the examples of a string in flat space
and the accelerated string in AdS space. We will then compare these
systems with the particle approximations.

\begin{figure}[t]
\begin{center}
\includegraphics[width=.47\textwidth]{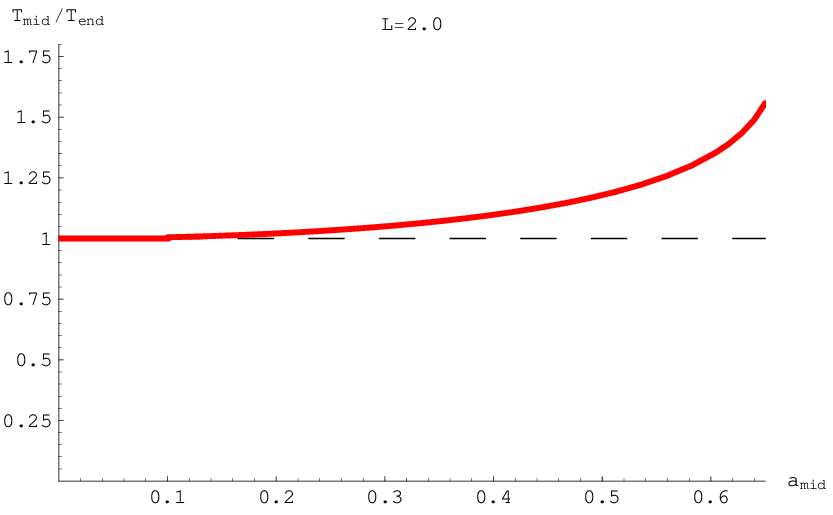}\quad
\includegraphics[width=.47\textwidth]{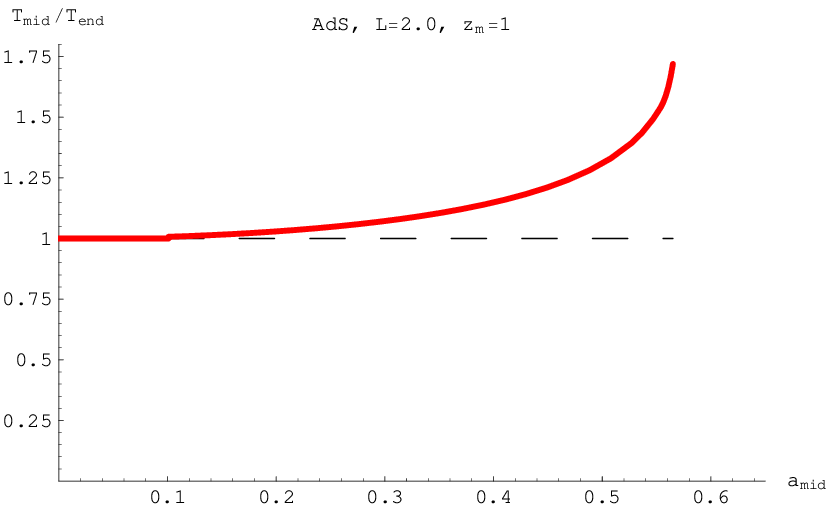}
\end{center}
\vspace{-3ex}
\caption{The ratio of the temperature felt by the string midpoint and
  the temperature felt by the endpoint, as a function of the
  endpoint acceleration. The figure on the left shows the result for
  a string in flat space, while the figure on the right shows the AdS
  case.  \label{f:Tmid_vs_aend}}
\end{figure}

Since the string is an extended object it is clear that if the string
does not move as a rigid object, different points on its worldvolume
could feel a different temperature. The explicit example of such a
behaviour is provided by the string in flat
space~\cite{Peeters:2007ti}.  Different points on the string move with
different (and constant) acceleration, with the maximal acceleration
(i.e.~temperature) experienced by the midpoint of the
string. Figure~\ref{f:Tmid_vs_aend} shows the dependence of the
midpoint temperature on the acceleration of the string endpoints. We
see that for small accelerations (when the string is almost straight
in Rindler space), the midpoint feels the same temperature as the
rest of the string. As the acceleration is increased, the temperature
of the midpoint starts to increase faster than the linear Unruh
relation between temperature and endpoint acceleration~\eqref{accelT}
(i.e.~what a point particle would experience). This holds true for
strings in flat space as well as for strings in AdS, but the increase
is more dramatic in the latter case.

\vfill\eject

%====================================================================
\section{Discussion and outlook}

In this paper we have analysed the effects of velocity and
acceleration on particles and strings, both in zero and finite
temperature backgrounds.

For the simplest case, in a background with zero Hawking temperature,
velocity has no effects. On the other hand, it is known that a
particle in flat space at constant acceleration will measure the Unruh
temperature~\cite{Unruh:1976db}.  The situation for strings is more
complicated. The AdS/CFT correspondence provides a perfect laboratory
in which one can study the effect of acceleration for a hadronic
string, as was shown in~\S~\ref{s:stringsAds}. We found, as expected,
that effects of acceleration are qualitatively similar to those of
temperature.  Generalising the previous analysis in flat
space~\cite{Frolov:1990ct,Peeters:2007ti}, we found that there exist
maximum accelerations for a string of given length or angular
momentum.  A surprise encountered in our analysis is that acceleration
is ``more efficient" than temperature in melting strings. This
statement means that, for a string of a given length, the Unruh
temperature of the string endpoints for which the string melts is
smaller than the Hawking temperature at which the static string would
undergo melting.  Qualitative similarities and quantitative
differences between temperature and acceleration for a non-rotating
string are summarised in figure~\ref{figcompare}.

When a finite temperature background is considered, we have seen that
velocity and acceleration produce \emph{independent} effects which
raise the observed temperature.  This is apparent
from~\eqref{e:Tvflat}, where $v\neq 0, \ a=0$ and~\eqref{T2T2T2} where
$v=0, \ a\neq0$. When both velocity and acceleration are present, the
situation is more complicated because the observed spectrum is far
from thermal. In~\cite{Liu:2006nn}, it was argued that the raise in
temperature due to velocity would enhance quarkonium suppression in a
quark-gluon-plasma. In view of our results, it is natural to expect
that if the highly energetic quarkonium state also feels a large
acceleration (or better said, deceleration) within the plasma, the
suppression would be enhanced further.

We see a number of directions for future study. One concerns the
analysis of the effect of rotation on observed temperature.  While we
have so far analysed the behaviour of linearly accelerated strings,
full-fledged holographic mesons also contain an acceleration component
due to the orbital motion. It is known that orbital motion itself does
\emph{not} lead to the observation of a thermal bath. However, the
spectrum of vacuum fluctuations \emph{is} to good approximation
thermal, and determines the response of a
detector~\cite{Letaw:1979wy}. For the superposed motion of angular and
linear acceleration, the Unruh effect has been studied in flat
space-time by~\cite{Letaw:1980ik,Letaw:1980ii}. Generically, for fixed
linear acceleration, the effective temperature of the vacuum
fluctuations is higher when there is a non-zero angular component. It
would be interesting to understand such computations in a holographic
context.

On a more fundamental level, it would be interesting to develop a
better understanding of the velocity dependence of temperature in the
context of general relativity. We have seen that the surface gravity
and GEMS methods both lead to a velocity dependence of the temperature
which is similar to the average of the angle-dependent temperature
obtained using other techniques. A proper understanding of this is
lacking, and requires a more detailed quantum-field theoretical
analysis, which might be feasible in a lower dimensional toy model.  We
hope to return to these questions in future work.

%====================================================================
\section*{Acknowledgements}

AP thanks Kevin Goldstein
for discussions.
The work of \mbox{AP \& KP} was supported in part by VIDI grant
016.069.313 from the Dutch Organisation for Scientific Research
(NWO). The work of AP is also partially supported by EU-RTN network
MRTN-CT-2004-005104 and INTAS contract 03-51-6346. KP thanks the
University of Durham for kind hospitality.

\vfill\eject
\appendix
%====================================================================
\section{Appendix: technical details}
\subsection{GEMS for AdS in global and Poincar\'e coordinates}
\label{a:GEMS_AdS}

The AdS space can be seen as a hyper-cylinder embedded in a
signature~(2,4) Minkowski space
\begin{equation}
X_0^2 - X_1^2 - X_2^2 - X_3^2 - X_4^2 + X_5^2 = R^2 \, . 
\end{equation}
A GEMS embedding of $\text{AdS}_5$ space in Poincar\'e
coordinates~\eqref{PoincareAds} into a signature (2,4) Minkowski space
is given by
\begin{equation}
\label{PoincareGEMS}
\begin{aligned}
X_0 &= \frac{1}{2u}\left(R^2+ 
\frac{u^2}{R^2} (R^2 + \sum_{i=1}^3 x_i^2 -t^2)\right)\,\,,
\\
X_i &=  \frac{u}{R}\,x_i   \,\qquad (i=1,2,3)\,\,,\nonumber\\
X_4 &= \frac{1}{2u}\left(R^2+ 
\frac{u^2}{R^2} ( - R^2 + \sum_{i=1}^3 x_i^2 -t^2)\right)\,\,,
\\
X_5 &= \frac{u}{R}\,t \,.
\end{aligned}
\end{equation}
A GEMS embedding of the $\text{AdS}_5$ space in global coordinates
is similarly given by
\begin{equation}
\begin{aligned}
X_0 &= R \cosh \rho\, \cos \tilde t \\
X_i &= R \sinh \rho\,  \omega_i \, ,  \quad \sum_i \omega_i^2  = 1 \, ,
\quad (i =1,2,3,4) \\
X_5 &= R \cosh \rho\, \sin \tilde t \, , 
\end{aligned}
\end{equation}

%--------------------------------------------------------------------
\subsection{GEMS for AdS-Schwarzschild black hole in Poincar\'e coordinates}
\label{a:GEMS_AdS_BH}

Let us consider a (3,5)-signature Minkowski space-time with signature
$(+,-,-,-,-,+,-,+)$.  We introduce
\begin{eqnarray}
X_0 &=& \frac{1}{2u}\left(R^2+ \frac{u^2}{R^2} (R^2 + \sum_{i=1}^3 x_i^2 )\right)\,\,,
\nonumber\\
X_i &=&  \frac{u}{R}\,x_i   \,\qquad (i=1,2,3)\,\,,\nonumber\\
X_4 &=& \frac{1}{2u}\left(R^2+ \frac{u^2}{R^2} ( - R^2 + \sum_{i=1}^3 x_i^2)\right) \,\,,
\nonumber\\
X_5 &=& \frac{R\,u}{2u_h}\,
\sqrt{1 - \frac{u_h^4}{u^4}}\,\sinh\left(\frac{2 u_h t}{R^2}\right)\,, \,\nonumber\\
X_6 &=& \frac{R\,u}{2u_h}\,
\sqrt{1 - \frac{u_h^4}{u^4}}\,\cosh\left(\frac{2 u_h t}{R^2}\right)\,, \,\nonumber\\
X_7 &=& \frac{R}{2} \int \sqrt{\frac{u^6+u^4 u_h^2 + 3u^2 u_h^4-u_h^6}
{u^6 u_h^2 + u^4 u_h^4}}{\rm d}u\,.
\label{PoincareBHGEMS2}
\end{eqnarray}
If we insert these expressions into
\begin{equation}
{\rm d}s^2 = {\rm d}X_0^2 - {\rm d}X_1^2- {\rm d}X_2^2- {\rm d}X_3^2-
{\rm d}X_4^2+ {\rm d}X_5^2 -{\rm d}X_6^2 +{\rm d}X_7^2\,,
\end{equation}
we recover the metric~\eqref{PoincareAdsBH}. It is not hard to check
that the GEMS computation using the above embedding yields the correct
observed temperature for a static observer (i.e.~sitting at
constant $x_1,x_2,x_3,u$) which is obtained by setting $v=0$ in
equation (\ref{Tforconstantv}).

\begin{small}
\setlength{\bibsep}{2.5pt}
%\bibliographystyle{kasper}
%\bibliography{kasbib}

\begingroup\raggedright\endgroup

\end{small}

\end{document}